\documentclass[11pt,a4paper]{article}
\begin{document}
\def\rf#1{(\ref{eq:#1})}
\def\lab#1{\label{eq:#1}}
\def\nonu{\nonumber}
\def\br{\begin{eqnarray}}
\def\er{\end{eqnarray}}
\def\be{\begin{equation}}
\def\ee{\end{equation}}
\def\lb{\lbrack}
\def\rb{\rbrack}
\def\Blb{\Bigl\lbrack}
\def\Brb{\Bigr\rbrack}
\def\lcurl{\left\{}
\def\rcurl{\right\}}
\def\>{\rangle}              %%  > for `ket'
\def\<{\langle}              %%  < for `bra'
\def\({\left(}
\def\){\right)}
\def\[{\left[}
\def\]{\right]}
\def\v{\vert}                     %% vertical bars
\def\bv{\bigm\vert}               %%
\def\tr{\mathop{\rm tr}}                  % tr - small trace
\def\Tr{\mathop{\rm Tr}}                  % Tr - big trace
\def\pr{\prime}
\def\ra{\rightarrow}
\def\lra{\longrightarrow}
\def\f{\frac}
\def\grad{\nabla}
\def\ti{\tilde}
\def\wti{\widetilde}
\def\eq{\!\!\!\! &=& \!\!\!\! }
% this paper
%\def\Ad{A^{\dagger}}
\def\p{\partial}
\def\jm{J_{-}}
\def\j3{J_3}
\def\Ad{{\cal A}^{\dagger}}
\def\A{{\cal A}}

\def\half{\frac{1}{2}}
\def\id{i\partial_\phi}
\def\hm{h_{min}}

%greek letters
\def\a{\alpha}
\def\b{\beta}
\def\c{\chi}
\def\t{\theta}
\def\T{\Theta}

\newcommand{\vp}{\varphi}
\newcommand{\sech} {{\rm sech}}
\newcommand{\cosech} {{\rm cosech}}
\newcommand{\psib} {\bar{\psi}}
\newcommand{\cosec} {{\rm cosec}}

\title{Study the spheroidal
wave functions by SUSYQM \footnote{ E-mail: tgh-2000@263.net,
tgh20080827@gmail.com, shuqzhong@gmail.com}}
\author{Guihua Tian$^{1,2}$,
\ Shuquan Zhong$^{1}$ \\
1.School of Science, Beijing University of Posts And
Telecommunications.
\\ Beijing 100876 China.\\
2.Department of Physics, University of Maryland, College Park, \\
Maryland 20742-4111 U.S.A}
\date{May 28, 2009}
\maketitle
\begin{abstract}
The perturbation method in supersymmetric quantum mechanics
(SUSYQM) is used to study the spheroidal wave functions'
eigenvalue problem. Expanding the super-potential in series of the
parameter $\a$, the first order term of ground eigen-value and the
eigen-function are gotten. In the paper, the very excellent
results are that all the first two terms approximation on
eigenfunctions obtained are in closed form. They give useful
information for the
involved physical problems in application of spheroidal wave functions. \\
\textbf{PACs:11.30Pb; 04.25Nx; 04.70-s}
\end{abstract}

\section*{Introduction}

  \ \ \ \ Since 1930s, spheroidal wave functions have made
strong contributions to extensively theoretical and practical
applications in pure mathematics, applied mathematics, physics and
engineering. They appear in the fields, such as wavelet, random
matrix, non-commute geometry, gravitational wave detection,
quantum field theory in curved space-time, black hole stable
problem; 3G mobile and broad band satellite telecommunication;
steady flow of a viscous fluid and so
on\cite{flammer}-\cite{slepian}. Nevertheless, they perhaps are
one of the hardest work for researchers. They welcome the new
thought and methods to deal with them. Since the appearance of the
supersymmetric quantum mechanics (SUSYQM), its great power of
solving the differential equation attracts tremendous attention.

In this paper, we first use SUSYQM to study The spheroidal
differential equations. First, brief introduction to the
spheroidal problems and the ordinary methods to treat them. The
spheroidal differential equations are
\begin{equation} \left[\frac{d}{dx}\left[(1-x^2) \frac{d}{d
x}\right]+E+ \a x^2
 -\frac{m^{2}}{1-x^2} \right]\Theta=0, x\in (-1,+1).\label{3}
\end{equation}
With the condition $\Theta $ is finite at the boundaries $x=\pm
1$, they consist of the Sturm-Liouville eigenvalue problem. The
parameter $E $ can only takes the values $E_0,\ E_1,\dots,
E_n,\dots$, which are called the eigenvalues of the problem, and
the corresponding solutions (the eigenfunctions) $\T_0,\T_1,\dots,
\T_n,\dots$   are called the spheroidal wave functions
\cite{flammer}-\cite{li}.

The equations (\ref{3}) have two parameters: $m$, $\a$. When
$\a=0$, the spheroidal wave functions reduce to the spherical wave
functions, that is, the associated Legendre-functions $P_l^m(x)$.
Though the spheroidal wave equations are extension of the ordinary
spherical wave functions equations, the difference between this
two kinds of wave equations are far greater than their
similarity\cite{flammer}. The spherical wave equations belong to
the case of the confluent super-geometrical equations with one
regular  and one irregular singularities, whereas the spheroidal
wave equations are the confluent Heun equations containing two
regular  and one irregular singularities. The extra singularity
makes it extremely difficult to solve
them\cite{flammer}-\cite{li}, so very little information have been
 obtained  concerning the analytically exact solutions. Therefore
approximate and numerical methods are two main resources to rely
on for the problem.

Traditionally, the three-term recurrence relation methods are used
to evaluate the eigenvalues and eigenfunctions of spheroidal wave
functions: one could solve the transcendental equation in
continued fraction form  or its equivalent or by power series
expansion etc\cite{flammer}-
\cite{li},\cite{caldwell}-\cite{beti}. For the details of these
methods and their advantage and disadvantage, one could see the
reference \cite{beti}. These methods mainly work for the numerical
purpose, and also rely heavily on the numerical method. However,
all previous works concentrate largely on the calculations of the
eigenvalues, and particularly emphasize the small parameter
approximation and large parameter limits form of the eigenvalues.
Little effort has been devoted to the related eigenfunctions due
to the difficulty and complexity.

Here, we give brief review on the eigenfunctions in small
parameter approximation. No matter what method may be used,  all
eigenfunctions come into this kind of form in the end \be
\Theta_{n}(x)=P_n^m+\sum_{q=1}^{\infty}\T_{nq}(x)\a^q\label
{small1}.\ee Though there are many excellent works on the
eigenvalues' approximation of small parameter, no good works exist
for the eigenfunctions approximation $\T_{nq}(x)$. Even for the
first eigenfunctions approximation, the existing works are only in
the series' form \be \T_{n1}(x)=\sum_{r=0,r\ne n
}^{\infty}B_{r}P_{m+r}^m(x)\label{small2}\ee with infinite numbers
$B_r$ needed to evaluate.  Obviously, this series form  does not
reveal much information about the eigen-functions, even about its
1st order approximation eigenfunctions $\T_{n1}(x)$ itself.

In recent years, supersymmetric quantum mechanics have attracted
tremendous attention for solvable potential problems. They not
only  provide clear insight into the factorization method of
Infeld and Hull \cite{Infeld}, but also greatly improve the
methods to solve the differential equations. See reference
\cite{Cooper} for review on its development.

The spheroidal eigenvalue problem is treated by the method of
SUSYQM. This is the first time for researchers to use SUSYQM to
study the ground eigenvalue and eigenfunction (that is $\T_0$) of
spheroidal wave functions in the small parameter $\a$
approximation.

In usual small parameter approximation method, the key concept is
the eigenfunctions and they are expanded in the form
(\ref{small1}). On the contrary, the super-potential is the
central concept in SUSYQM, and is expanded in the series form of
the parameter $\a$. This new method is applied to study the
spheroidal equations and unexpected results are obtained: the
ground eigenfunction of the first order is in closed form, this in
turn gives useful information on the eigenfunction and is helpful
for their application. We also get the ground function for higher
order terms in parameter $\alpha$.

\section*{The ground eigenvalue and eigenfunction in the first order }
\ \ \ \

In the following, we will use the new perturbation method in
supersymmetry quantum to resolve the spheroidal eigenfunctions'
problem.

Though the form (\ref{3}) is more familiar for research, the
problem is easier to solve in the original differential equation
than in the equation (\ref{3}). The original form is obtained from
the eq.(\ref{3}) by the transformation \be x=\cos\t, \ee that is,
\begin{equation}
\left[\frac{1}{\sin \theta}\frac{d}{d\theta}\left(\sin \theta
\frac{d}{d \theta}\right)+ \a \cos ^2 \theta -\frac{m^{2}}{\sin ^2
\theta}\right]\Theta=-E\Theta\label{2}
\end{equation}
the corresponding  boundary conditions become $\T$ is finite at
$\theta=0,\ \pi$,

 One writes the eqn.(\ref{2}) in the form of the
Schr$\ddot{o}$dinger equation by the transformation
\begin{equation}
\Theta =\frac{\Psi}{\sin^{\frac12} \theta}\label{transform}
\end{equation}
the differential equations turn out to be
\begin{eqnarray}
\frac{d^2\Psi}{d\theta^2}+\left[\frac14+ \a\cos ^2 \theta
 -\frac{m^{2}-\frac14}{\sin ^2
\theta}+E \right]\Psi=0\label{main eq}
\end{eqnarray}
and the boundary conditions become
\begin{equation}
\Psi|_{\theta=0}=\Psi|_{\theta=\pi}=0
\end{equation}
From the equation (\ref{main eq}), one knows the potential is
\begin{equation}
V(\theta,\a, m)=-\frac14- \a\cos ^2 \theta
+\frac{m^{2}-\frac14}{\sin ^2 \theta}\label{potential m s b}
\end{equation}
The super-potential $W$ is determined by the potential
$V(\t,\a,m)$ through the Reccita's equation \begin{equation}
W^2-W'=V(\theta,\alpha)-E_0\label{potential and w relation}
\end{equation} where the substraction of the ground energy just
makes the eqn.(\ref{main eq}) factorable. Actually this equation
is the same hard to treat as that in the original form (\ref{main
eq}). The approximate method naturally comes to one's mind. Hence,
when the absolute value of $\a$ is small, it is the the
super-potential $W$ that could be expanded as series of the
parameter $\a$, that is,
\begin{equation}
W=W_0+\alpha W_1+\alpha ^2 W_2+\alpha ^3 W_3+\ldots .\label{W}
\end{equation}

\begin{eqnarray}
 W^2&-&W'= W_0^2-W'_0+\alpha \left(2W_0W_1-W'_1\right)+\alpha ^2
\left(2W_0W_2+W_1^2-W'_2\right)\nonumber\\
&+& \alpha ^3 \left(2W_0W_3+2W_1W_2-W'_3\right)+\alpha ^4
\left(2W_0W_4+2W_1W_3+W_2^2-W'_4\right)+\ldots .\label{V-W1}
\end{eqnarray}

One can write the perturbation equation as
\begin{equation}
W^2-W'=V(\theta,\alpha,
m)-\sum_{n=0}^{\infty}2E_{0n}\alpha^n=-\frac14+\frac{m^2-\f14}{\sin
^2 \theta}- \alpha\cos ^2 \theta
-\sum_{n=0}^{\infty}2E_{0n}\alpha^n\label{2potential m=s=0 alpha}
\end{equation}
There are two lower indices in the parameter $E_{0n}$ with the
index $0$ refereing to the ground state and the other index $n$
meaning the nth term in parameter $\a$. The last term
$\sum_{n=0}^{\infty}2E_{on}\alpha^n$ is subtracted from the above
equation in order to make the ground state energy actually zero
for the application of the theory of SUSYQM. Later, one must add
the term to our calculated eigen-energy.  Comparing the equations
(\ref{V-W1}), (\ref{potential m s b}), and (\ref{2potential m=s=0
alpha}), one could get
\begin{eqnarray}
&&W_0^2-W'_0=-\frac14+\frac{m^2-\f14}{\sin ^2 \theta}-2E_{00}\label{0-term}\\
&&2W_0W_1-W'_1=- \alpha\cos ^2 \theta-2E_{01}\label{1-term}\\
&&2W_0W_2+W_1^2-W'_2=-2E_{02}\label{2-term}\\
&&2W_0W_3+2W_1W_2-W'_3=-2E_{03}\label{3-term}\\
&&2W_0W_4+2W_1W_3+W_2^2-W'_4=-2E_{04}\label{4-term}\\
&&\ \ \ \ \ \ \ \ \vdots
\end{eqnarray}
From the eq.(\ref{0-term}), we get
\begin{equation}
W_0=-\(m+\f12\) \cot\theta,\ 2E_{00}=m(m+1).\label{w-0}
\end{equation}
Then, we can write the other equations more concisely
\begin{eqnarray}
&&W'_1+(2m+1)\cot\theta W_1=\cos ^2 \theta+2E_{01}\label{1-1-term}\\
&&W'_2+(2m+1)\cot\theta W_2= W_1^2+2E_{02}\label{2-1-term}\\
&&W'_3+(2m+1)\cot\theta W_3=2W_1W_2+2E_{03}\label{3-1-term}\\
&&W'_4+(2m+1)\cot\theta W_4=2W_1W_3+W_2^2+2E_{04}\label{4-1-term}\\
&&W'_5+(2m+1)\cot\theta W_5=2W_1W_4+2W_2W_3+2E_{05}\label{5-1-term}\\
&&\ \ \ \ \ \ \ \ \vdots
\end{eqnarray}
After obtaining the zero term $W_0$ for the super-potential $W$,
the first order $W_1$ can be gotten as
\begin{equation}
W_1=\frac{\bar{A}_1}{\sin^{2m+1} \theta}
\end{equation}
with\br \f{d \bar{A_1}}{d \t}&=& \sin^{2m+1}\t \(\cos^2\t+2E_{01}\)\\
\bar{A_1}&=&\int \(\sin^{2m}\t \cos^2\t+2E_{01}\sin^{2m}\t\)\sin\t
d\t \er Suitably changing the independent variable to $x=\cos\t $
and expanding the term as following \br \sin^{2m}\t
&=&\(1-\cos^2\t\)^{m}=\(1-x^2\)^{m}=\sum_{k=0}^{m}(-1)^k{m\choose
k}x^{2k}\er where ${m\choose k}=\f{m\times (m-1)\times \dots
\times (m-k+1)}{k!}$, then it reaches
\begin{equation}
\bar{A}_1=\sum_{k=0}^{m}(-1)^{k+1}{m\choose
k}\[\f{\cos^{2k+3}\t}{2k+3}+\f{2E_{01}\cos^{2k+1}\t}{2k+1}\].
\end{equation}
 So,
\begin{equation}
W_1=\frac{\bar{A}_1}{\sin^{2m+1}\theta}=
\frac{\sum_{k=0}^{m}(-1)^{k+1}{m\choose
k}\[\f{\cos^{2k+3}\t}{2k+3}
+\f{2E_{01}\cos^{2k+1}\t}{2k+1}\]}{\sin^{2m+1}\theta}\label{w1 in
complex }.
\end{equation}
 The quantity $E_{01}$ needs to be determined by the boundary
 conditions. This in turn require to calculate the ground eigenfunction upon to the first order
  by
\br \Psi_0&=&N\exp\[-\int Wd\theta\]\\&=& N\exp\[-\int W_0d\theta\
-\alpha \int
 W_1d\theta\]*\exp{O(\a^2)}\\
&=& N\sin^{m+\f12}\t \exp\[-\alpha \int
 W_1d\theta\]*\exp{O(\a^2)}\label{f-w relation}. \er
Whenever the eigenfunction is obtained, the boundary conditions
$\Psi|_{\theta=0}=\Psi|_{\theta=\pi}=0$ would choose the proper
$E_{01}$. This sounds very easy, but is a tough task in reality.
The complete calculating process is left to the appendix 1. The
results are \br
&&2E_{01}=-\f1{2m+3}\label{e1 in simple }\\
&&  W_1=\f{\sin\t\cos\t}{2m+3}\label{good result of w1}.\er

With the first order term of the super-potential $W_1$, one could
compute the second term $W_2$ by the same process. This is not
easy either. The appendix 2 gives the detail of the calculation.
Nevertheless, the results are elegant: \br
&&2E_{02}=-\f{2m+2}{(2m+3)^3(2m+5)}\label{e2 in simple }\\
&& W_2=
\[\f{-\sin\t
\cos\t}{(2m+3)^3(2m+5)}+\f{\sin^3\t \cos\t}{(2m+3)^2(2m+5)}\]
\label{w2 in complex 1}.\er

The ground eigenfunction upon to the second order becomes \br
\Psi_0&=&N\exp\[-\int Wd\theta\]\\&=& N\exp\[-\int W_0d\theta\
-\alpha \int
 W_1d\theta-\a^2\int W_2d\t\]*\exp{O(\a^3)}\\
&=&\left(sin{\theta}\right)^{m+\frac12 } \exp{\[-\frac{\alpha
\sin^2\theta}{4m+ 6}\]}\nonumber\\ &*&  \exp{\[\f{\a^2\sin^2\t
}{2(2m+3)^3(2m+5)}-\f{\a^2\sin^4\t
}{4(2m+3)^2(2m+5)}\]}*\exp{O(\a^3)} .\er

When $m=0$, These results (\ref{e1 in simple }),(\ref{good result
of w1}),(\ref{e2 in simple }),(\ref{w2 in complex 1}) reduce
respectively to \br && 2E_{01}=-\f13,\ W_{1}=\f13\sin\t\cos\t;\\
&& 2E_{02}=-\f2{135},\ W_2=-\f1{135}\sin\t \cos\t +\f1{45}\sin^3\t
\cos\t. \er

 They are in complete accordance with that in reference \cite{tian} where
 the spheroidal wave functions are treated by SUSYQM in the case
 $m=0$.

Back  to equation (\ref{3}) with  their relationship
(\ref{transform}), we  could obtain \br \Theta_0 &=&
\left(1-x^2\right)^{\f
m2}\exp{\left(-\frac{\alpha\(1-x^2\)}{4m+6}\right)}\nonumber\\ &*&
\exp{\[\f{\a^2(1-x^2) }{2(2m+3)^3(2m+5)}-\f{\a^2(1-x^2)^2
}{4(2m+3)^2(2m+5)}\]}*\exp{O(\a^3)}\label{good ground 1 }.\er

Expanding the exponential functions in the above equation, the
results elegantly turn out as \br \Theta_0 &=&
\left(1-x^2\right)^{\f m2}\[1-\a \frac{1-x^2}{4m+6}+\a^2\(\f{1-x^2
}{2(2m+3)^3(2m+5)}+\f{(1-x^2)^2 }{8(2m+3)(2m+5)}\)\]\nonumber \\
&& +0(\a^3) \label{good ground 2 }.\er
 The above equation clearly shows how
the ground state function changes as the function of variable $x$
when $\a$ is small and could be compared to the non-perturbation
case $P_m^m=(1-x^2)^{\f m2}$. The result is much better than the
usual result of the series form \be \Theta_0(x)=P_m^m+\a
\sum_{q=1}^{\infty}B_qP_{m+q}^m(x)+\a^2
\sum_{q=1}^{\infty}C_qP_{m+q}^m(x)+O(\a^3)\ee with infinite
numbers $B_q,\ q=1,\ 2,\ \dots\ $ need to be determined for the
first order term and $C_q,\ q=1,\ 2,\ \dots\ $ need to be
determined for the second order term. From another point view, the
results in the eqns.(\ref{good ground 1 }), (\ref{good ground 2 })
give the method to determine the infinite numbers $B_q,\ C_q,\
q=1,\ 2,\ \dots\ $.

In conclusion, SUSYQM provides a new opportunity to treat the
spheroidal wave functions and indeed they give new results in the
eqns. (\ref{good ground 1 }),(\ref{good ground 2 }). Further
calculations can be done by the same way, nevertheless, the higher
order term $W_n$ is more complex than the lower one. The maximus
of $W_1$ and $W_2$ satisfy \br \max{W_1}=\f16,\
\max{W_2}=\f{\sqrt{11}}{2160}<\f1{540}<\max{W_1}. \er
 Though the further calculation is not processed here, the
 reasonable guess is that the higher term $W_n$ is, the smaller
 its maximus. The guess mainly comes from the calculation of the
 quantity $W_1,\ W_2$, see the appendix 1 and appendix 2 for
 details. If the guess is right, \br
 W(\t)=\sum_{n=0}^{\infty}W_{n}\a^n \er might be analytic function
 in all complex plane $\a$, the only singularity of the function $W$ as the variable $\a$ is $\a=\infty$.
  Only the calculation in the appendixes can
 process on and on, could the guess be tested. Might some day the
 computer can do the work, this is the reason that the calculations in
 appendixes are extremely detailed.

\section*{Acknowledgements}
We greatly appreciate the hospitality of the Gravitation Theory
group and the MCFP in UMD . This work was supported in part by the
National Science Foundation of China under grants No.10875018,
No.10773002.

\section*{Appendix1: Simplification of the super potential of the first term $W_1$}
\begin{appendix}
\ \ \ \ The calculations are very complex, we rewrite the
eq.(\ref{w1 in complex }) here again for convenience
\begin{equation}
W_1=\frac{\bar{A}_1}{\sin^{2m+1}\theta}=
\frac{\sum_{k=0}^{m}(-1)^{k+1}{m\choose
k}\[\f{\cos^{2k+3}\t}{2k+3}+\f{2E_{01}\cos^{2k+1}\t}{2k+1}\]}{\sin^{2m+1}\theta}\label{w1
in complex 2 }.
\end{equation}
As state before, the quantity $E_{01}$ is determined by the
requirement that the eigenfunction is zero at the boundaries
$\t=0,\ \pi$; this in turn demands $\int W_1d\t$ finite at the
boundary.
  Therefore, the calculation of the
term $\int W_1d\t$ is first processed . By transformation \be
\tau=\sin\t\label{trans2}\ee and denoting $\int W_1d\t$ by $I$ ,
it reads \br I=\int W_1d\t &=& \int
\frac{\sum_{k=0}^{m}(-1)^{k+1}{m\choose
k}\[\f{(1-\tau^2)^{k+1}}{2k+3}+\f{2E_{01}(1-\tau^2)^{k}}{2k+1}\]}{\tau^{2m+1}}d\tau\nonumber\er
Using formula \br
&&  \(1-\tau^2\)^{k}=\sum_{l=0}^{k}(-1)^l{k\choose l}\tau^{2l}\\
&&\(1-\tau^2\)^{k+1}=\sum_{l=0}^{k+1}(-1)^l{k+1\choose
l}\tau^{2l}\label{expanding (a+b) },\er it becomes \br I&=&  \int
\sum_{k=0}^{m}(-1)^{k+1}{m\choose
k}\[\f{\sum_{l=0}^{k+1}(-1)^l{k+1\choose l}\tau^{2l-2m-1}}{2k+3}
+\f{2E_{01}\sum_{l=0}^{k}(-1)^l{k\choose l}\tau^{2l-2m-1  }}{2k+1}\]d\tau\nonumber\\
. &=& \sum_{k=0}^{m}(-1)^{k+1}{m\choose
k}\[\sum_{l=0}^{k+1}\f{(-1)^l{k+1\choose
l}\tau^{2l-2m}}{(2k+3)(2l-2m)}+\sum_{l=0}^{k}
\f{2E_{01}(-1)^l{k\choose l}\tau^{2l-2m}}{(2k+1)(2l-2m)}\]\\ &=&
\sum_{k=0}^{m}\sum_{l=0}^{k+1}\f{(-1)^{k+l+1}{m\choose
k}{k+1\choose
l}\tau^{2l-2m}}{(2k+3)(2l-2m)}+\sum_{k=0}^{m}\sum_{l=0}^{k}
\f{2E_{01}(-1)^{k+l+1}{m\choose k}{k\choose l}\tau^{2l-2m
}}{(2k+1)(2l-2m)}\er In order to exchange the sums order in the
above equations, one must notice the fact that $l\le k+1\
\Rightarrow k\ge l-1$ in the first term
$\sum_{k=0}^{m}\sum_{l=0}^{k+1}\f{(-1)^{k+l+1}{m\choose
k}{k+1\choose l}\tau^{2l-2m}}{(2k+3)(2l-2m)}$ and $l\le k\
\Rightarrow k\ge l$ in the second term
$\sum_{k=0}^{m}\sum_{l=0}^{k} \f{2E_{01}(-1)^{k+l+1}{m\choose
k}{k\choose l}\tau^{2l-2m }}{(2k+1)(2l-2m)}$. So \br I &=&
\sum_{l=0}^{m+1}\sum_{k=l-1\ge 0 }^{m}\f{(-1)^{k+l+1}{m\choose
k}{k+1\choose
l}\tau^{2l-2m}}{(2k+3)(2l-2m)}+\sum_{l=0}^{m}\sum_{k=l}^{m}
\f{2E_{01}(-1)^{k+l+1}{m\choose k}{k\choose l}\tau^{2l-2m  }}{(2k+1)(2l-2m)}\\
&=& \sum_{l=0}^{m}\[\sum_{k=l-1\ge 0 }^{m}\f{(-1)^{k+l+1}{m\choose
k}{k+1\choose l}}{(2k+3)(2l-2m)}+2E_{01}\sum_{k=l}^{m}
\f{(-1)^{k+l+1}{m\choose k}{k\choose l}}{(2k+1)(2l-2m)}
\]\tau^{2l-2m}+\f{\tau^2}{4m+6}\er where the first sum under the
condition $l=m+1$ becomes $\f{\tau^2}{4m+6}$.

Defining\br N_{1\ l}&=&\sum_{k=l-1\ge 0
}^{m}\f{(-1)^{k+l+1}{m\choose k}{k+1\choose l}}{(2k+3)(2l-2m)},\ l=0,1,2,\dots,\label{n1}\\
 N_{2\ l}&=& \sum_{k=l}^{m}
\f{(-1)^{k+l+1}{m\choose k}{k\choose l}}{(2k+1)(2l-2m)},\
l=0,1,2,\dots \label{n2}\er it gets
 \br I&=& \sum_{l=0}^{m}
\[N_{1\ l}+2E_{01}N_{2\ l}\]\tau^{2l-2m}+\f{\tau^2}{4m+6}\label{sum of I}.\er

There are terms $\tau^{2l-2m}$ in the above equation. With the
eq.(\ref{trans2}) and the fact $l< m$, these term as
$\sin^{2l-2m}\t$ become infinite as $\t \ra 0,\ \pi$. By
eq.(\ref{f-w relation}) and the eigenfunction's boundary condition
at $\t=0,\ \pi$, the coefficients of those terms must be zero.
There is only one quantity $E_{01}$ unfixed, could  one choose
proper $E_{01}$ to to make the eigenfunction finite at the
boundaries? Actually, one only has one choice to select $E_{01}$
by \br N_{1\ 0}+2E_{01}N_{2\ 0}=0,\ \ N_{1\ 0}=\f1{2m}\sum_{k=0
}^{m}\f{(-1)^{k+2}{m\choose k}}{(2k+3)},\ \
 N_{2\ 0}= \f1{2m}\sum_{k=0}^{m}
\f{(-1)^{k+2}{m\choose k}}{(2k+1)}\label{choice E01},\er do the
other terms in eqn.(\ref{sum of I}) automatically become zero
under the condition (\ref{choice E01} ).

Fortunately, this can be done and the following is the proof. The
inductive reasoning is used to give the proof. In order to
determine the quantity $E_{01}$ under the condition (\ref{choice
E01}), one must first simplify $N_{1\ 0},\ N_{2\ 0}$. From the
formula ( see reference \cite{grad} on page 389) \br &&
\int_{0}^{1}(1-\tau^2)^md\tau= \int_0^{\t=\f{\pi}2}\cos^{2m+1}\t
d\t = \f{(2m)!!}{(2m+1)!!}, \er and the similarly one\br
&& \int_{0}^{1}\tau^2(1-\tau^2)^md\tau=\int_{0}^{1}(1-\tau^2)^md\tau-\int_{0}^{1}(1-\tau^2)^{m+1}d\tau\\
&& =\[\f{(2m)!!}{(2m+1)!!}-\f{(2m+2)!!}{(2m+3)!!}\]\er and also
with the formula (\ref{expanding (a+b) }), it is easy to obtain
\br &&
\int_{0}^{1}\tau^2(1-\tau^2)^md\tau=\int_0^1\sum_{k=0}^m{m\choose k}(-1)^k\tau^{2k+2}d\tau=\sum_{k=0}^m\f{(-1)^{k+2}{m\choose k}}{2k+3}\\
&& \int_{0}^{1}(1-\tau^2)^md\tau=\int_0^1\sum_{k=0}^m{m\choose
k}(-1)^k\tau^{2k}d\tau=\sum_{k=0}^m\f{(-1)^{k+2}{m\choose
k}}{2k+1}\er Comparing these equations with that of (\ref{choice
E01}), one reaches \br N_{1\ 0}&=&\f1{2m}\sum_{k=0
}^{m}\f{(-1)^{k+2}{m\choose k}}{2k+3}=\f1{2m}\[\f{(2m)!!}{(2m+1)!!}-\f{(2m+2)!!}{(2m+3)!!}\]\nonumber\\
N_{2\ 0}&=& \f1{2m}\sum_{k=0}^{m} \f{(-1)^{k+1}{m\choose
k}}{2k+1}=\f1{2m}\f{(2m)!!}{(2m+1)!!}.\er Consequently, the
quantity $2E_{01}$ is simplified as \br 2E_{01}=-\f{N_{1\
0}}{N_{2\ 0}}=-\f1{2m+3}\label{choice E01-2}.\er

$N_{1\ 0}+2E_{01}N_{2\ 0}=0$ is guaranteed  by the choice in
eqn.(\ref{choice E01-2}). According to inductive reasoning, one
needs to prove  \br M_{1\  l+1}=N_{1\ l+1}+2E_{01}N_{2\ l+1}=0 \er
under the assumption that \br M_{1\  l}=N_{1\ l}+2E_{01}N_{2\
l}=0.\er

The key idea is to find the connection between  the terms $M_{1\
l}=N_{1\ l}+2E_{01}N_{2\ l}$ and $M_{1\ l+1}=N_{1\
l+1}+2E_{01}N_{2\ l+1}$. By the definitions of $N_{1\ l },\ N_{2\
l},\  l=0,\ 1,\ 2,\ \dots $ in eqns.(\ref{n1})-(\ref{n2}), one has
\br N_{1\ l+1}=&=&\sum_{k=l
}^{m}\f{(-1)^{k+l+2}{m\choose k}{k+1\choose l+1}}{(2k+3)(2l+2-2m)},\\
 N_{2\ l+1}&=& \sum_{k=l+1}^{m}
\f{(-1)^{k+l+2}{m\choose k}{k\choose l+1}}{(2k+1)(2l+2-2m)}.\er
Due to the following relation \br
 \f1{2k+3}{k+1\choose l+1}&=&\f{(k+1)!}{(2k+3)(k-l)!(l+1)!}=\f{(k+1)!}{(2k+3)(k+1-l)!l!}*\f{k+1-l}{(l+1)}\nonumber \\
&=&\f{(k+1)!}{(k+1-l)!l!}
*\f{k+\f32-(l+\f12)}{(l+1)(2k+3)}\nonumber
\\&=&-\f{l+\f12}{l+1}\times
\f1{2k+3}{k+1\choose l}+\f1{2l+2}{k+1\choose l},\label{choose
relation}\er it is easy to get \br (2l+2-2m)N_{1\
l+1}&=&\sum_{k=l }^{m}\f{(-1)^{k+l}{m\choose k}{k+1\choose l+1}}{2k+3}\\
&=&\sum_{k=l }^{m}(-1)^{k+l}{m\choose k}\[-\f{l+\f12}{l+1}\times
\f1{2k+3}{k+1\choose l}+\f1{2l+2}{k+1\choose l}\]\\&=&
\sum_{k=l-1\ge 0 }^{m}\[\f{l+\f12}{l+1}\times
\f1{2k+3}\](-1)^{k+l+1}{m\choose k}{k+1\choose l}-\f{{m\choose
l-1}}{2(l+1)}\nonumber\\&+& \sum_{k=l
}^{m}\f{(-1)^{k+l}}{2l+2}{m\choose k}{k+1\choose l}. \er Now, the
calculation becomes \br (2l+2-2m)N_{1\
l+1}=\f{l+\f12}{(l+1)}(2l-2m)N_{1\ l}-\f{{m\choose
l-1}}{2(l+1)}+\sum_{k=l }^{m}(-1)^{k+l}{m\choose
k}\f1{2l+2}{k+1\choose l}\label{relation of n1-1}\er under with
the help of eqn.(\ref{n1}). Definition of the quantity \br Q_{1\
l}&=&-\f{{m\choose l-1}}{2l+2}+\sum_{k=l
}^{m}\f{(-1)^{k+l}{m\choose k}{k+1\choose l}}{2l+2}\nonumber\\
&=&\f1{2l+2}\sum_{k=l-1 }^{m}(-1)^{k+l}{m\choose k}{k+1\choose l},
\label{Q1 result}\er may make the equation (\ref{relation of
n1-1}) simple as \br (2l+2-2m)N_{1\
l+1}=\f{l+\f12}{(l+1)}(2l-2m)N_{1\ l}+Q_{1\ l }\ l=0,\ 1,\ 2,\
\dots,\  m-1.\label{relation of n1}\er

In completely similar way, one could get the relation between
$N_{2,\ l}$ and $N_{2\ l+1}$. the formula \br
\f1{2k+1}{k\choose l+1}&=& \f1{2k+1}\f{k!}{(k-l-1)!(l+1)!}=\f1{2k+1}\f{k!}{(k-l)!(l)!}*\f{k-l}{l+1}\nonumber \\
&=&\f{k!}{(k-l)!(l)!}*\f{(k+\f12)-(l+\f12)}{(2k+1)(l+1)}\nonumber \\
&=& -\f{l+\f12}{l+1}\times \f1{2k+1}{k\choose
l}+\f1{2l+2}{k\choose l}\er helps to simplify  the quantity $N_{2\
l+1 }$

\br (2l+2-2m)N_{2\  l+1}&=&
\sum_{k=l+1}^{m} \f{(-1)^{k+l}{m\choose k}{k\choose l+1}}{2k+1}\\
&=&\sum_{k=l+1}^{m} (-1)^{k+l}{m\choose k}\[-\f{l+\f12}{l+1}\times
\f1{2k+1}{k\choose l}+\f1{2l+2}{k\choose l}\]\\&=&\sum_{k=l}^{m}
\f{l+\f12}{l+1}\times \f{(-1)^{k+l+1}}{2k+1}{m\choose k}{k\choose
l}-\f{l+\f12}{l+1}\times
\f{(-1)^{2l+1}}{2l+1}{m\choose l}{l\choose l} \nonumber\\
&+&\sum_{k=l+1}^{m} \f{(-1)^{k+l}}{2l+2}{m\choose k}{k\choose
l}.\er Using the eqn.(\ref{n2}) and the fact
$-\f{l+\f12}{l+1}\times \f{(-1)^{2l+1}}{2l+1}{m\choose l}{l\choose
l}=\f1{2l+2}{m\choose l}$, it is easy to obtain \br (2l+2-2m)N_{2\
l+1}&=&\f{(l+\f12)(2l-2m)}{(l+1)}N_{2\ l}+\f{{m\choose
l}}{2l+2}+\sum_{k=l+1}^{m} \f{(-1)^{k+l}}{2l+2}{m\choose
k}{k\choose l}\label{relation of n2-1 }.\er The similar definition
of the quantity \br Q_{2\ l}&=&\f{{m\choose
l}}{2l+2}+\sum_{k=l+1}^{m} \f{(-1)^{k+l}}{2l+2}{m\choose
k}{k\choose l}\nonumber \\ &=& \f{1}{2l+2}\sum_{k=l}^{m}
(-1)^{k+l}{m\choose k}{k\choose l}\label{Q2 result} \er makes the
equation (\ref{relation of n2-1 }) simply become \br
(2l+2-2m)N_{2\ l+1}&=&\f{(l+\f12)(2l-2m)}{(l+1)}N_{2\ l}+Q_{2\
l},\ l=0,\ 1,\ 2,\ \dots,\  m-1.\label{relation of n2 }\er

The eqns.(\ref{relation of n1}),(\ref{relation of n2 }) tell that
\br M_{1\ l+1}&=&N_{1\ l+1}+2E_{01}N_{2\
l+l}\\&=&\f{(l+\f12)(2l-2m)}{(l+1)(2l-2m+2)}M_{1\  l} +\[Q_{1,\ l
}+2E_{01}Q_{2\ l}\],\ l=0,\ 1,\ 2,\ \dots,\  m-1\label{M relation
}.\er the relation between $M_{1\ l}$,$M_{1\ l+1}$ would be the
desired result if one could prove that $Q_{1\ l}=Q_{2\ l}=0,\
l=0,\ 1,\ 2,\ \dots,\  m-1 $. This is the following work.

 Using the formula \br
(fg)^{(m)}=\sum_{k=0}^{m}{m\choose k}f^{(m-k)}g ^{(k)},\er under
the special case of $f=\tau,\ g=(1-\tau)^{m}$, one gets \br
\[\tau(1-\tau)^m\]^{(l)}&=&\tau\[(1-\tau)^m\]^{(l)}+{l\choose 1}\[(1-\tau)^m\]^{(l-1)}\\
&=&(-1)^lm(m-1)\dots
(m-l+1)\tau(1-\tau)^{m-l}\nonumber\\&+&(-1)^{l-1}l\times
m(m-1)\dots (m-l+2)(1-\tau)^{m-l+1}\label{useful q1-1}.\er  The
calculation may also go by different way: \br
\[\tau(1-\tau)^m\]^{(l)}&=&\[\sum_{k=0}^{m}(-1)^k{m\choose k}\tau^{k+1}\]^{(l)}\\&=& \sum_{k=l-1}^{m}(-1)^k{m\choose k}(k+1)k(k-1)\dots
(k-l+2)\tau^{k-l+1}\\
&=&\f1{(l-1)!}\sum_{k=l-1}^{m}(-1)^k{m\choose k}{k+1\choose
l}\tau^{k-l+1}\label{useful q1-2 }\er

Comparing the results of eqns. (\ref{useful q1-1}), (\ref{useful
q1-2 }) and taking the function $\[\tau(1-\tau)^m\]^{(l)}$ at
$\tau=1$ under the condition $l<m$, the good result reaches \br
\[\sum_{k=l-1}^{m}(-1)^k{m\choose k}{k+1\choose l}\tau^{k-l+1}\]_{\tau=1}=\sum_{k=l-1}^{m}(-1)^k{m\choose k}{k+1\choose l}=0\label{useful
q1-3 }.\er By eqn.(\ref{Q1 result}), it is easy to get \br
Q_{1l}=0,\ l=0,\ 1,\ 2,\ \dots,\  m-1 .\label{q1=0}\er Similarly,
one also could prove that \br Q_{2l}=0,\ l=0,\ 1,\ 2,\ \dots,\ m-1
.\label{q2=0}\er Therefore, the eqns.(\ref{M relation }),
(\ref{q1=0}), (\ref{q2=0}) imply that\br M_{1\  l+1}=0,\ l=0,\ 1,\
2,\ \dots,\  m-1 \er under the condition \br M_l=0;\er by
induction one gets \br M_{n}=0,\ n=1,2,\dots ,  m.\er Hence, the
boundary conditions could be satisfied by just selecting the only
one quantity $2E_{01}=-\f1{2m+3}$. With the good results $M_{1\
l}=N_{1\ l}+2E_{01}N_{2\ l}=0, l=0,\ 1,\ 2,\ \dots,\ m$ , one can
greatly simplify the first order super-potential $W_1$ in the
eqn.(\ref{w1 in complex 2 }). By similar method as before, rewrite
$W_1$ by changing the independent variable to \be \tau=\sin\t \ee
and expanding terms $(1-\tau^2)^{k+1},\ (1-\tau^2)^k$, that is \br
&&W_1=\frac{\sum_{k=0}^{m}(-1)^{k+1}{m\choose k}\[\f{(1-\tau^2)^{k+\f32}}{2k+3}+\f{2E_{01}(1-\tau^2)^{k+\f12}}{2k+1}\]}{\tau^{2m+1}}\\
&=&(1-\tau^2)^{\f12} \sum_{k=0}^{m}(-1)^{k+1}{m\choose
k}\[\f{\sum_{l=0}^{k+1}(-1)^l{k+1\choose l}}{2k+3}
+\f{2E_{01}\sum_{l=0}^{k}(-1)^l{k\choose
l}}{2k+1}\]\tau^{2l-2m-1}\er then changing the order of the sums
just as before and dividing $W_1$ by $(1-\tau^2)^{\f12}$ for good
looking in the formula, it reads \br \f{W_1}{(1-\tau^2)^{\f12}}&=&
\sum_{l=0}^{m+1}\sum_{k=l-1\ge 0
}^{m}\f{(-1)^{k+l+1}{m\choose k}{k+1\choose l}\tau^{2l-2m-1}}{(2k+3)}\nonumber\\
&+&\sum_{l=0}^{m}\sum_{k=l}^{m}
\f{2E_{01}(-1)^{k+l+1}{m\choose k}{k\choose l}\tau^{2l-2m-1  }}{(2k+1)}\\
&=& \sum_{l=0}^{m}\sum_{k=l-1\ge 0 }^{m}\f{(-1)^{k+l+1}{m\choose
k}{k+1\choose l}}{(2k+3)}\tau^{2l-2m-1}\nonumber\\
&+&\[\f{(-1)^{k+l+1}{m\choose k}{k+1\choose
l}}{(2k+3)}\tau^{2l-2m-1}\]_{{l=m+1}\atop{k=m}}+2E_{01}\sum_{k=l}^{m}
\f{(-1)^{k+l+1}{m\choose k}{k\choose l}}{(2k+1)} \tau^{2l-2m-1}\\
&=& \sum_{l=0}^{m}\[\sum_{k=l-1\ge 0 }^{m}\f{(-1)^{k+l+1}{m\choose
k}{k+1\choose l}}{(2k+3)}+ 2E_{01}\sum_{k=l}^{m}
\f{(-1)^{k+l+1}{m\choose k}{k\choose l}}{(2k+1)}\] \tau^{2l-2m-1}\nonumber\\&+&\f{\tau}{2m+3}\\
&=&
\[N_{1\ l}+2E_{01}N_{2\ l}\]*(2l-2m)\tau^{2l-2m-1}+\f{\tau}{2m+3}\\&=& \f{\tau}{2m+3}\er
So the quantity $W_1$ could be written tidily as \br
W_1=\f{\tau(1-\tau^2)^{\f12}}{2m+3}=\f{\sin\t \cos\t}{2m+3}.\er

\section*{Appendix2: Simplification of the super potential of the first term $W_2$}
\ \ \ \ Though the process of the calculation of $W_2$ repeats
that of $W_1$ in the appendix 1, there still are some needs to
write it down. Perhaps this may be used for further computation,
even by computer.

\br W_2&=& \f{\bar{A_2}}{\sin^{2m+1}\t}\er where \br
\bar{A_2}&=& \int (W_1^2 +2E_{02})\sin^{2m+1}\t d\t\\
&=&\int \[\f1{(2m+3)^2}\sin^{2m+3}\t \cos^2\t+2E_{02}\sin^{2m+1}\t\] d\t\\
&=&-\int  \[\f1{(2m+3)^2}(1-x^2)^{m+1}
x^2+2E_{02}(1-x^2)^{m}\]dx\ \ \ \dots, x=\cos\t \\
&=&\int  \[\f1{(2m+3)^2}\sum_{k=0}^{m+1}(-1)^{k+1}{m+1\choose
k}x^{2k+2}+2E_{02}\sum_{k=0}^{m}(-1)^{k+1}{m\choose
k}x^{2k}\]dx \\
&=& \sum_{k=0}^{m+1}\f{(-1)^{k+1}{m+1\choose
k}x^{2k+3}}{(2m+3)^2(2k+3)}+2E_{02}\sum_{k=0}^{m}\f{(-1)^{k+1}{m\choose
k}x^{2k+1}}{2k+1} \\&=&\sum_{k=0}^{m+1}\f{(-1)^{k+1}{m+1\choose
k}\cos^{2k+3}\t}{(2m+3)^2(2k+3)}+2E_{02}\sum_{k=0}^{m}\f{(-1)^{k+1}{m\choose
k}\cos^{2k+1}\t}{2k+1}\er The other important term is \br
&&II=\int W_2d\t\\&&=\int
\[\sum_{k=0}^{m+1}\f{(-1)^{k+1}{m+1\choose
k}\cos^{2k+3}\t}{(2m+3)^2(2k+3)\sin^{2m+1}\t}+2E_{02}\sum_{k=0}^{m}\f{(-1)^{k+1}{m\choose
k}\cos^{2k+1}\t}{(2k+1)\sin^{2m+1}\t}\]d\t. \er By the
transformation $\tau=\sin\t$, one may have \br &&\int W_2d\t =
\int
\[\sum_{k=0}^{m+1}\f{(-1)^{k+1}{m+1\choose
k}(1-\tau^2)^{k+1}}{(2m+3)^2(2k+3)\tau^{2m+1}}+2E_{02}\sum_{k=0}^{m}\f{(-1)^{k+1}{m\choose
k}(1-\tau^2)^{k}}{(2k+1)\tau^{2m+1}}\]d\tau
\\&&=\sum_{k=0}^{m+1}\sum_{l=0}^{k+1}\f{(-1)^{k+l+1}{m+1\choose
k}{k+1\choose l
}\tau^{2l-2m}}{(2m+3)^2(2k+3)(2l-2m)}+2E_{02}\sum_{k=0}^{m}\sum_{l=0}^{k}\f{(-1)^{k+l+1}{m\choose
k}{k\choose l }\tau^{2l-2m}}{(2k+1)(2l-2m)}\er

Exchanging the sum order, it reads \br &&
II=\sum_{l=0}^{m+2}\sum_{k=l-1\ge 0
}^{k+1}\f{(-1)^{k+l+1}{m+1\choose k}{k+1\choose l
}\tau^{2l-2m}}{(2m+3)^2(2k+3)(2l-2m)}+2E_{02}\sum_{l=0}^{m}\sum_{k=l}^{m}\f{(-1)^{k+l+1}{m\choose
k}{k\choose l }\tau^{2l-2m}}{(2k+1)(2l-2m)}\\
&&= \sum_{l=0}^{m}\sum_{k=l-1\ge 0
}^{m+1}\f{(-1)^{k+l+1}{m+1\choose k}{k+1\choose l
}\tau^{2l-2m}}{(2m+3)^2(2k+3)(2l-2m)}+2E_{02}\sum_{l=0}^{m}\sum_{k=l}^{m}\f{(-1)^{k+l+1}{m\choose
k}{k\choose l
}\tau^{2l-2m}}{(2k+1)(2l-2m)}\\&&+\sum_{l=m+1}^{m+2}\sum_{k=l-1\ge
0 }^{m+1}\f{(-1)^{k+l+1}{m+1\choose k}{k+1\choose l
}\tau^{2l-2m}}{(2m+3)^2(2k+3)(2l-2m)}\\ &&
=\sum_{l=0}^{m}\[\sum_{k=l-1\ge 0
}^{m+1}\f{(-1)^{k+l+1}{m+1\choose k}{k+1\choose l
}\tau^{2l-2m}}{(2m+3)^2(2k+3)(2l-2m)}+2E_{02}\sum_{k=l}^{m}\f{(-1)^{k+l+1}{m\choose
k}{k\choose l
}\tau^{2l-2m}}{(2k+1)(2l-2m)}\]\\&&+\sum_{l=m+1}^{m+2}\sum_{k=l-1\ge
0 }^{m+1}\f{(-1)^{k+l+1}{m+1\choose k}{k+1\choose l
}\tau^{2l-2m}}{(2m+3)^2(2k+3)(2l-2m)}\er

It is better to define the following quantities \br N_{3, \
l}=\sum_{k=l-1\ge 0 }^{m+1}\f{(-1)^{k+l+1}{m+1\choose
k}{k+1\choose l }}{(2m+3)^2(2k+3)(2l-2m)},\ l=0,\ 1,\ 2,\ \dots,\
m\label{n3}\er \br \ N_{4\
l}=\sum_{k=l}^{m}\f{(-1)^{k+l+1}{m\choose k}{k\choose l
}}{(2k+1)(2l-2m)}, \ l=0,\ 1,\ 2,\ \dots,\ m\label{n4}\er and
simplify the last term \br && \sum_{l=m+1}^{m+2}\sum_{k=l-1\ge 0
}^{m+1}\f{(-1)^{k+l+1}{m+1\choose k}{k+1\choose l
}\tau^{2l-2m}}{(2m+3)^2(2k+3)(2l-2m)}\\&&=-\f{(2m+6)\tau}{(2m+3)^3(2m+5)}+\f{\tau^3}{3(2m+3)^2(2m+7)}.\er
Then the express becomes \br II=\sum_{l=0}^{m}\[N_{3\
l}+2E_{02}N_{4\
l}\]\tau^{2l-2m}-\f{(2m+6)\tau}{(2m+3)^3(2m+5)}+\f{\tau^3}{3(2m+3)^2(2m+7)}\er
Just as done before, one needs to select proper $E_{02}$ to make
all $N_{3\ l}+2E_{02}N_{4\ l}$ zero. By comparing the quantities
$N_{1\ l},N_{2\ l},N_{3\ l},N_{4\ l}$, the good relations among
them could be revealed: \br N_{4\ l}=N_{2\ l},\ l=0,\ 1,\ 2,\
\dots,\ m\er and the great similarity between $N_{1\ l},N_{3\ l}$.
The transformation forms between $N_{1\ l+1},N_{1\ l}$, $N_{3\
l+1},N_{3\ l}$ are similar.

\br N_{3, \ l+1}=\sum_{k=l\ge 0 }^{m+1}\f{(-1)^{k+l+2}{m+1\choose
k}{k+1\choose l+1 }}{(2m+3)^2(2k+3)(2l-2m+2)}\er By the relation
(\ref{choose relation}) \br
 \f1{2k+3}{k+1\choose l+1}=-\f{l+\f12}{l+1}\times
\f1{2k+3}{k+1\choose l}+\f1{2l+2}{k+1\choose l}\er

one could get \br && (2m+3)^2(2l+2-2m)N_{3\
l+1}=\sum_{k=l }^{m+1}\f{(-1)^{k+l}{m+1\choose k}{k+1\choose l+1}}{2k+3}\\
&=&\sum_{k=l }^{m+1}(-1)^{k+l}{m+1\choose
k}\[-\f{l+\f12}{l+1}\times \f1{2k+3}{k+1\choose
l}+\f1{2l+2}{k+1\choose l}\]\\&=& \sum_{k=l-1\ge 0
}^{m+1}\[\f{l+\f12}{l+1}\times \f1{2k+3}\](-1)^{k+l+1}{m+1\choose
k}{k+1\choose l}-\f{{m+1\choose l-1}}{2(l+1)}\nonumber\\&+&
\sum_{k=l }^{m+1}\f{(-1)^{k+l}}{2l+2}{m+1\choose k}{k+1\choose l};
\er with the help of eqn.(\ref{n3}), the calculation becomes \br
N_{3\ l+1}=\f{(l+\f12)(2l-2m)}{(l+1)(2l+2-2m)}N_{3\
l}+\f{\[-\f{{m+1\choose l-1}}{2(l+1)}+\sum_{k=l
}^{m+1}(-1)^{k+l}\f{{m+1\choose k}{k+1\choose
l}}{2l+2}\]}{(2m+3)^2(2l+2-2m)}.\label{relation of n3-1}\er The
definition of the quantity \br Q_{3\ l}&=&-\f{{m+1\choose
l-1}}{2l+2}+\sum_{k=l
}^{m+1}\f{(-1)^{k+l}{m+1\choose k}{k+1\choose l}}{2l+2}\nonumber\\
&=&\f1{2l+2}\sum_{k=l-1 }^{m+1}(-1)^{k+l}{m+1\choose k}{k+1\choose
l}\label{Q3 result}\er makes the above equation (\ref{relation of
n3-1}) become \br N_{3\
l+1}=\f{(l+\f12)(2l-2m)}{(l+1)(2l+2-2m)}N_{3\ l}+\f{Q_{3\
l}}{(2m+3)^2(2l+2-2m)},\ l=0,\ 1,\ 2,\,\dots,\ m-1.\label{relation
of n3}.\er Another useful relation is \br N_{4\ l+1}&=&N_{2\
l+1}=\f{(l+\f12)(2l-2m)}{(l+1)(2l+2-2m)}N_{2\
l}\\&=&\f{(l+\f12)(2l-2m)}{(l+1)(2l+2-2m)}N_{4\ l}\er by the use
of $Q_{2\ l}=0$. Now the quantities $M_{2\ l}$ is defined as  \br
M_{2\ l}=N_{3\ l}+2E_{02}N_{4\ l},\ l=0,\ 1,\ 2,\ \dots,\ m\er
then the relation between $M_{2\ l}$ and $M_{2\ l+1}$ is \br M_{2\
l+1} =\f{(l+\f12)(2l-2m)}{(l+1)(2l-2m+2)}M_{2\ l}+\f{Q_{3\
l}}{(2m+3)^2(2l+2-2m)},l=0,1, \dots, m-1\label{M2 relation }.\er

In the following, it will be proven that $Q_{3\ l}=0$. With the
help \br (fg)^{(m)}=\sum_{k=0}^{m}{m\choose k}f^{(m-k)}g ^{(k)}\er
under the special case $f=\tau,\ g=(1-\tau)^{m+1}$, one gets \br
\[\tau(1-\tau)^{m+1}\]^{(l)}&=&\tau\[(1-\tau)^{m+1}\]^{(l)}+{l\choose 1}\[(1-\tau)^{m+1}\]^{(l-1)}\\
&=&(-1)^lm(m+1)\dots
(m-l+2)\tau(1-\tau)^{m-l+1}\nonumber\\&+&(-1)^{l-1}l\times
m(m+1)\dots (m-l+2)(1-\tau)^{m-l+2}\label{useful q3-1}.\er The
alternative way to compute is \br
\[\tau(1-\tau)^{m+1}\]^{(l)}&=&\[\sum_{k=0}^{m+1}(-1)^k{m+1\choose k}\tau^{k+1}\]^{(l)}\\&=& \sum_{k=l-1}^{m+1}(-1)^k{m+1\choose k}(k+1)k(k-1)\dots
(k-l+2)\tau^{k-l+1}\\
&=&\f1{(l-1)!}\sum_{k=l-1}^{m+1}(-1)^k{m+1\choose k}{k+1\choose
l}\tau^{k-l+1}.\label{useful q3-2 }\er

The valuation of  $\[\tau(1-\tau)^{m+1}\]^{(l)}$ at $\tau=1$ under
the condition $l<m+1$ is  \br
\[\sum_{k=l-1}^{m+1}(-1)^k{m+1\choose k}{k+1\choose l}\tau^{k-l+1}\]_{\tau=1}=\sum_{k=l-1}^{m+1}(-1)^k{m+1\choose k}{k+1\choose l}=0\label{useful
q3-3 }.\er From the eqn.(\ref{Q3 result}), it is easy to get \br
Q_{3\ l}=0\ l=0,\ 1,\ 2,\,\dots,\ m-1.\label{q3=0}\er

Finally, one gets the good result \br M_{2\ l+1}
&=&\f{(l+\f12)(2l-2m)}{(l+1)(2l-2m+2)}M_{2\ l},\ l=0,\ 1,\
2,\,\dots,\ m-1. \label{M2 relation2 }\er So the choice of $M_{2\
0}=0$ guarantees $M_{2,\ l}=1,\ 2,\,\dots,\ m$. The quantity
$E_{02}$ is obtained by \br E_{02}=-\f{N_{3\ 0}}{N_{4\ 0}}.\er As
in the case $N_{1\ 0}$, \br N_{3\ 0}&=&\f1{2m}\sum_{k=0
}^{m+1}\f{(-1)^{k+2}{m+1\choose k}}{(2m+3)^2(2k+3)}=\f1{2m(2m+3)^2}\[\f{(2m+2)!!}{(2m+3)!!}-\f{(2m+4)!!}{(2m+5)!!}\]\nonumber\\
N_{4\ 0}&=&N_{2\ 0}= \f1{2m}\sum_{k=0}^{m} \f{(-1)^{k+2}{m\choose
k}}{2k+1}=\f1{2m}\f{(2m)!!}{(2m+1)!!}.\er Hence \br
E_{02}=-\f{N_{3\ 0}}{N_{4\ 0}}=-\f{2m+2}{(2m+3)^3(2m+5)}.\er  The
expression  \br W_2=\f{\sum_{k=0}^{m+1}\f{(-1)^{k+1}{m+1\choose
k}\cos^{2k+3}\t}{(2m+3)^2(2k+3)}+2E_{02}\sum_{k=0}^{m}\f{(-1)^{k+1}{m\choose
k}\cos^{2k+1}\t}{2k+1}}{\sin^{2m+1}\t}\er could greatly simplified
by $\tau=\sin\t$ and the use of the elegant formula $M_{2\ l}=0$:
\br && \f{W_2}{\cos\t}=\sum_{k=0}^{m+1}\f{(-1)^{k+1}{m+1\choose
k}(1-\tau^2)^{k+1}}{(2m+3)^2(2k+3)\tau^{2m+1}}+2E_{02}\sum_{k=0}^{m}\f{(-1)^{k+1}{m\choose
k}(1-\tau^2)^{k}}{(2k+1)\tau^{2m+1}}\\&&=\sum_{k=0}^{m+1}\sum_{l=0}^{k+1}\f{(-1)^{k+l+1}{m+1\choose
k}{k+1\choose l
}}{(2m+3)^2(2k+3)\tau^{2m-2l+1}}+2E_{02}\sum_{k=0}^{m}\sum_{l=0}^{k}\f{(-1)^{k+l+1}{m\choose
k}{k\choose l}}{(2k+1)\tau^{2m-2l+1}}.\er Exchanging the sums
order, it is easy to get \br &&
\f{W_2}{\cos\t}=\sum_{l=0}^{m+2}\sum_{k=l-1\ge 0
}^{m+1}\f{(-1)^{k+l+1}{m+1\choose k}{k+1\choose l
}}{(2m+3)^2(2k+3)\tau^{2m-2l+1}}+2E_{02}\sum_{l=0}^{m}\sum_{k=l}^{m}\f{(-1)^{k+l+1}{m\choose
k}{k\choose l}}{(2k+1)\tau^{2m-2l+1}},\er comparing the
eqns.(\ref{n3}) and (\ref{n4}), one may have \br &&
\f{W_2}{\cos\t}=\f{\sum_{l=0}^{m}\[N_{3\ l}+2E_{02}N_{4\
l}\]}{\tau^{2m-2l+1}}+\sum_{l=m+1}^{m+2}\sum_{k=l-1\ge 0
}^{m+1}\f{(-1)^{k+l+1}{m+1\choose k}{k+1\choose l
}}{(2m+3)^2(2k+3)\tau^{2m-2l+1}}\\
&& =\sum_{l=m+1}^{m+2}\sum_{k=l-1\ge 0
}^{m+1}\f{(-1)^{k+l+1}{m+1\choose k}{k+1\choose l
}}{(2m+3)^2(2k+3)\tau^{2m-2l+1}}\\ &&=
\f{-\tau}{(2m+3)^3(2m+5)}+\f{\tau^3}{(2m+3)^2(2m+5)}\er by the use
of $M_{2\ l}=N_{3\ l}+2E_{02}N_{4\ l}=0$. The elegant form of
$W_2$ is \br
W_{2}=\[\f{-1}{(2m+3)^3(2m+5)}+\f{\sin^2\t}{(2m+3)^2(2m+5)}\]\sin\t
\cos\t .\er This ends the appendix 2.

\end{appendix}

\end{document}